%% file: QSseg_arxiv.tex
\documentclass[article,twocolumn,showpacs,aps,pre,floatfix,superscriptaddress,longbibliography]{revtex4-1}
\usepackage{graphicx,subfigure,epsfig,verbatim,psfrag,amsmath,amssymb,color}
\usepackage{indentfirst}
\usepackage{textcomp}
\usepackage{latexsym}
\usepackage{amssymb}
\usepackage{amsmath}
\usepackage{lipsum}
\usepackage{soul}
\usepackage{xcolor}
\usepackage{color}
\usepackage{epstopdf}
\usepackage{placeins}
\usepackage{float}
\usepackage{amsmath}

\usepackage{xcolor}
\usepackage[urlcolor=blue]{hyperref}      
\hypersetup{
    colorlinks = true,                    
    citecolor = {blue},
    linkcolor = {purple},
           }
\newcommand{\be}{\begin{equation}}
\newcommand{\ee}{\end{equation}}
\newcommand{\bea}{\begin{eqnarray}}
\newcommand{\eea}{\end{eqnarray}}

\begin{document}	
\title{Visual configuration segmentation of quantum states for phase identification in many-body systems}
\author{Yuan Yang}
\affiliation{School of Physical Sciences, University of Chinese Academy of Sciences, P. O. Box 4588, Beijing 100049, China}
\author{Zhengchuan Wang}
\email[Corresponding author. ] {Email: wangzc@ucas.ac.cn}
\affiliation{School of Physical Sciences, University of Chinese Academy of Sciences, P. O. Box 4588, Beijing 100049, China}
\author{Shi-Ju Ran}\email[Corresponding author. ] {Email: sjran@cnu.edu.cn}
\affiliation{Department of Physics, Capital Normal University, Beijing 100048, China}
\author{Gang Su}
\email[Corresponding author. ] {Email: gsu@ucas.ac.cn}
\affiliation{Kavli Institute for Theoretical Sciences, and CAS Center for Excellence in Topological Quantum Computation, University of Chinese Academy of Sciences, Beijing 100190, China}
\affiliation{School of Physical Sciences, University of Chinese Academy of Sciences, P. O. Box 4588, Beijing 100049, China}
\date{\today}
\begin{abstract}
Artificial intelligence provides an unprecedented perspective for studying phases of matter in condensed-matter systems.
Image segmentation is a basic technique of computer vision that belongs to a branch of artificial intelligence. In this work, we propose a segmentation scheme named visual configuration segmentation (VCS) to unveil quantum phases and quantum phase
transitions in many-body systems. By encoding the information of renormalized quantum states into a color image and segmenting the color image through the VCS, the renormalized quantum states can be visualized, from which quantum phase transitions can be
revealed and the corresponding critical points can be identified.
Our scheme is benchmarked on several strongly correlated spin systems, which does not depend on the priori knowledge of order parameters of quantum phases.
This demonstrates the potential to disclose the underlying structure of quantum phases by the techniques of computer vision.
\end{abstract}
\maketitle

\section{Introduction}
In recent years, artificial intelligence~(AI)~\cite{turing1950i, silver2016mastering, rebentrost2014quantum} has caused a great impact on physics. One of the main branches of AI is the computer vision~\cite{ marr2010vision, sonka1993image},
where computers or machines can gain high-level understanding from digital images or videos and
implement the tasks that the human visual system can do ~\cite{marr2010vision, sonka1993image}.
Until now, vision is still the most important way for human's brain to obtain and analyze the information~\cite{marr2010vision,yao2017quantum}. Based on the rapid development of computer vision, various fields
including economics, biomedicine, face recognition and automatic pilot have been significantly advanced~\cite{gonzalez2009digital,
lake2015human, jean2016combining}.

One of the fundamental techniques of computer vision is the image segmentation~\cite{jaiswal2013a}, which
can divide the digital image into multiple regions according to some homogeneity criterion.
Image segmentation is the first step in many attempts to analyze or interpret an image automatically~\cite{singh2010a}.
The simplest and efficient method of image segmentation is the thresholding method \cite{shapiro2000}.
According the threshold, an image can be transformed to a binary image.
Proper binarization of the images is very important for separating the
foreground object from the background.

\begin{figure}[htb]
\includegraphics[width=0.95 \columnwidth]{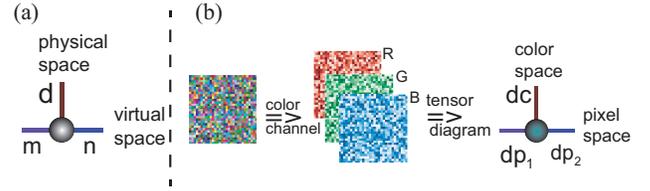}
\caption{(Color online) (a) A 3-index tensor related to matrix product states (MPS), where one unit of MPS is shown.
$d$ is physical index with dimension $\sigma$, which represents the local physical space of particles. $m$ and $n$ are the virtual indexes with dimension $\chi_1$ and $\chi_2$, which are related to the entanglement between two subsystems.
(b) The RGB color image with random pixels, which consist of three color channels that R, G, B correspond to red, green, blue colors. Each color channel is related to a pixel matrix. So, RGB color image can also be represented a by 3-index tensor,
where $d_c$ is the color space, $dp_1$ and $dp_2$ are the pixel space.}
\label{fig::map_image}
\end{figure}

In the same period of AI, the classical simulation of
quantum many-body system has also made significant progresses based on the numerical
renormalization group methods.
From the density matrix renormalization group (DMRG)~\cite{white1992density} to
tensor networks (TNs)~\cite{schollwock2011density,VC06MPSFaithfully,ran2020tensor}, one can simulate the
ground states of quantum many-body systems approaching to the thermodynamic limit.
For a local tensor $T = \{T^{d=1,\cdots,\sigma}_{m=1,\cdots,\chi_1; n=1,\cdots,\chi_2}\}$ in TNs
such as the matrix product states \cite{VC06MPSFaithfully}, it can be equivalently treated as a
color image as the physical index $d$ can be interpreted as the color channel, as shown in Fig.~\ref{fig::map_image},
where each channel is related to a gray image with size $\chi_1 \times \chi_2$ \cite{kottmann2020unsupervised}.

In this work, we design a visualization scheme of matrix product states (MPS) based image segmentation scheme \cite{stockman2001computer}
named visual configuration segmentation (VCS) to reveal the quantum phases and phase transitions.
Based on this scheme, renormalized quantum states (the centre tensor of MPS by DMRG) are visualized by binary images.
The selection of thresholds is the key for image segmentation \cite{jaiswal2013a,singh2010a},
which has a direct influence on the final results of image processing
such as the edge detection \cite{marr1980theory}.
The traditional way for the binarization of image is first to gray the image and
then to choose a threshold \cite{otsu1979a,niblack1986introduction}.
Different from the traditional way for binarization,
in our scheme VCS, we first choose a threshold in each channel to get binary images, then
we calculate the absolute differences of each two binary images to get the finally
binary images (see \emph{Methods}).
When the renormalized quantum states are in the same phase, the corresponding binary images look similar. When
there is a phase transition, the textures of the corresponding binary images exhibit sharp differences.

Our scheme is benchmarked on one-dimensional (1D) quantum lattice models,
where various quantum phases including
those within and beyond Landau paradigm.
Different from the conventional approaches in many-body physics
that often require the information of order parameters
to character the quantum phases and quantum phase transitions,
our scheme shows an intuitive sense of vision of the renormalized quantum states by image segmentation technique, and
reveals the quantum phase transitions directly by visual sensing \cite{khatami2020visualizing}.
Our work paves a new way to study quantum many-body systems by revealing the quantum phases and quantum phase transitions
through the computer vision techniques.

\section{Methods}
For a three-index tensor $T = \{T_{m=1,\cdots,\chi_1;n=1,\cdots,\chi_2 }^{d=1,\cdots,\sigma}\}$, $T_{m,n}^d$ is the tensor elements, and the dimensions of the three indexes $m$, $n$, and $d$ are denoted as $\chi_1$, $\chi_2$ and $\sigma$. $T$ can be interpreted as a color image through a map $\mathcal{G}:$ $T = \{T_{m,n}^d\}
\overset{\mathcal{G}} {\to} T^{\mathcal{G}}=\{T_{m,n}^{\mathcal{G},d}\}$, where $T_{m,n}^{\mathcal{G},d}$ represents the pixel
value at the position $(m,n)$ in the color channel $d$. Each channel yields a gray image with the size $\chi_1\times \chi_2$.

\begin{figure}[htb]
\includegraphics[width=0.95 \columnwidth]{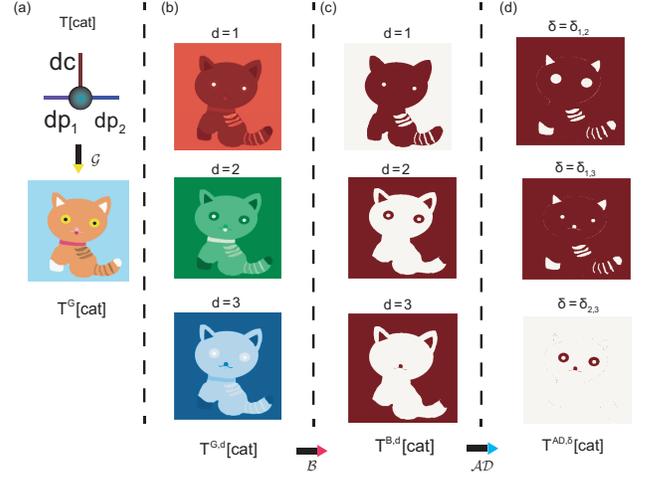}
\caption{(Color online) Graphically illustrate the image segmentation scheme VCS.
(a) The color image $T^{\mathcal{G}}[cat]$, which is the visualization of 3-index tensor $T[cat]$ by mapping $\mathcal{G}$,
where $\mathcal{G} = \mathcal{I}$ is the identical transformation.
(b) The gray images $T^{\mathcal{G},d}[cat]$ with $d=1$, $2$, and $3$ are corresponding to three color channels $R$ (red), $G$ (green), and $B$ (blue).
(c) The binary images $T^{\mathcal{B},d}[cat]$ corresponding to three color channels $R$, $G$, and $B$ are obtained by mapping $\mathcal{B}$.
(d) The binary images $T^{\mathcal{AD},\delta}$
are obtained by mapping $\mathcal{AD}$ according to Eq.~(\ref{eq::AD}).}
\label{fig::cat_image}
\end{figure}

Based on the threshold method of image segmentation \cite{stockman2001computer,jaiswal2013a,singh2010a},
we design a scheme named VCS to segment image $T^{\mathcal{G}}$.
We take a cat image $T^{\mathcal{G}}[cat]$ (i.e. the image interpretation of tensor $T[cat]$ shows a shape of cat,) as an instance to illustrate VCS (Fig.~\ref{fig::cat_image}).

There are two main steps to complete VCS.
The first step is the binarization in the color channels, i.e. each gray image $T^{\mathcal{G},d}[cat]$
is mapped onto a binary image by the map $\mathcal{B}$ as $T^{\mathcal{G},d} \overset{\mathcal{B}} {\to} T^{\mathcal{B},d}$
satisfying
\begin{equation}
{T}_{m,n}^{\mathcal{B},d} =\left\{
\begin{array}{rcl}
1   &      & {T_{m,n}^{\mathcal{G},d} \geq \alpha^d}\\
\\
0   &      & {T_{m,n}^{\mathcal{G},d} < \alpha^d}\\
\end{array} \right.
\label{eq::B}
\end{equation}
where $\alpha^d$ is the threshold value for $d$-th color channel.
We choose the mean value of pixels as threshold value $\alpha^d$ i.e. $\alpha^d = \frac{1}{\chi_1\times \chi_2}\sum_{m,n}T_{m,n}^{\mathcal{G},d}[cat]$.
The binary cat image $T^{\mathcal{B},d}[cat]$ is shown in Fig.~\ref{fig::cat_image} (c),
where three binary images corresponding to channels $R$ $(d=1)$, $G$ $(d=2)$, and $B$ $(d=3)$
that contain the original cat.

The second step is to obtain the final binary image
$T^{\mathcal{AD}, \delta_{d_1,d_2} }$ by the map $\mathcal{AD}$ that the absolute difference between
any two binary images $T^{\mathcal{G},d_1}, T^{\mathcal{G},d_2} \overset{\mathcal{AD}}{\to} T^{\mathcal{AD},\delta_{d_1,d_2}}$ :
\begin{equation}
T^{\mathcal{AD},\delta_{d_1, d_2}} = |T^{\mathcal{B}, d_1} - T^{\mathcal{B}, d_2}|,
\label{eq::AD}
\end{equation}
with $d_1\neq d_2$ and $T^{\mathcal{AD},\delta_{d_1, d_2}} = T^{\mathcal{AD},\delta_{d_2, d_1}}$.
The number of the final binary images is $\frac{\sigma (\sigma -1 )}{2}$.
For the cat image $T^{\mathcal{G}}[cat]$, three binary images $T^{\mathcal{AD}, \delta_{1,2}}[cat]$, $T^{\mathcal{AD}, \delta_{1,3}}[cat]$ and $T^{\mathcal{AD}, \delta_{2,3}}[cat]$ are
shown in Fig.~\ref{fig::cat_image} (d).

Both kinds of binary images $T^{\mathcal{B}}$ and $T^{\mathcal{AD}}$ are the segmentation of original colorful image $T^{\mathcal{G}}$.
However, the concentrated information of the $T^{\mathcal{G}}$ by
map $\mathcal{B}$ and map $\mathcal{AD}$ are different.
Taking the cat image $T^{\mathcal{G}}[cat]$ as example,
$T^{\mathcal{B}}[cat]$ show the whole outlines of the cat in Fig.~\ref{fig::cat_image} (c) while $T^{\mathcal{AD}}[cat]$
highlights the local key details of cat such as the eyes and ears in Fig.~\ref{fig::cat_image} (d).
Our scheme manages to segment the details of the cat from the backgrounds.

One shall note that in the existing ways \cite{otsu1979a},
the segmentation of images will be finished after the step of mapping $\mathcal{B}$.
However, in our scheme, we make a further binary segmentation by the map $\mathcal{AD}$.

\begin{figure*}[htb]
\includegraphics[width=1.8 \columnwidth]{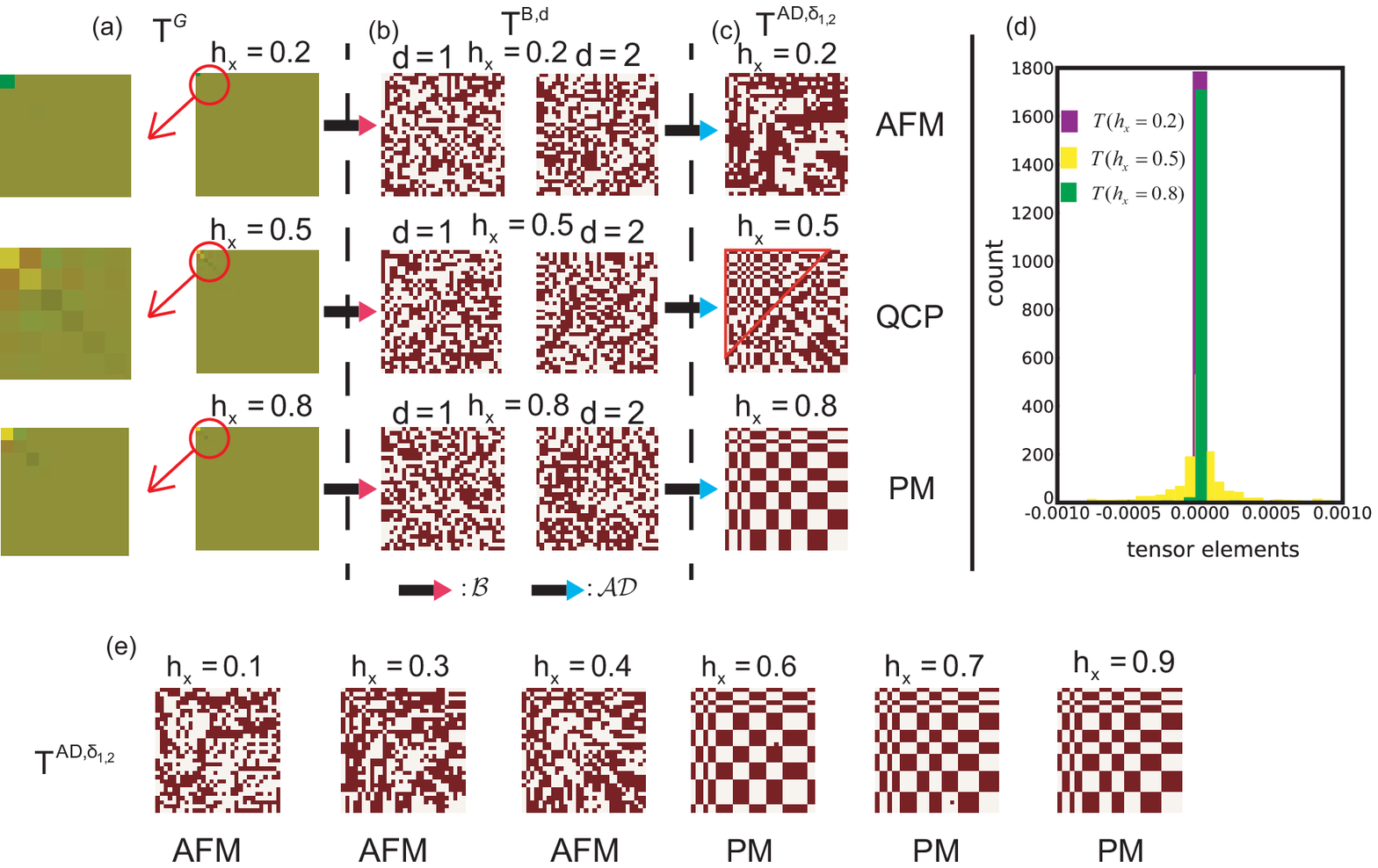}
\caption{(Color online) The color image $\Phi^{\mathcal{G}}$, binary images $\Phi^{\mathcal{B}}$ and binary images $\Phi^{\mathcal{AD}, \delta_{1,2}}$ of the center tensor $\Phi$ of the ground states of the transverse field Ising model (TFIM) with different transverse fields $h_x$ = $0.2$, $0.5$, and $0.8$ are in (a), (b) and (c), respectively.
(a) The distinguished details of $\Phi^{\mathcal{G}}$ with different $h_x$ are marked by read circles.
(b) The binary images $\Phi^{\mathcal{B},d=1~(2)}$ with different $h_x$, and its textures are always non-regular.
(c) The antiferromagnetic (AFM) and paramagnetic (PM) phases are clearly specified by binary images $\Phi^{\mathcal{AD},\delta_{1,2}}$.
The partial regular texture of $\Phi^{\mathcal{AD},\delta_{1,2}}$ when the system at QCP ($h_x=0.5$) is marked by red triangular.
(d) is the histogram that counts the occurring times of the tensor elements of $\Phi$ with $h_x$ = $0.2$, $0.5$, and $0.8$.
(e) More binary images $\Phi^{\mathcal{AD},\delta_{1,2}}$ with different transverse fields that $h_x=0.1$, $0.3$, $0.4$, $0.6$, $0.7$
and $0.9$. For DMRG, we take the size of the system L = 80 and the dimension cut-off $\chi$ = 30.
The size of binary images $\Phi^{\mathcal{AD},\delta_{1,2}}$ is $\chi\times\chi$.}
\label{fig::Ising_image}
\end{figure*}

\section{Entropy and correlation from binary images}
We introduce two concepts that are information entropy and correlation to qualitatively characterize the segmentation.
In particular, information entropy introduced by Shannon\cite{shannon1948a,papoulis2002probability} allows us to measure the amount of the information given by the distribution of binary pixel values. The pixels of binary images have two values $0$ and $1$, from which the Shannon entropy is calculated by $S=-p(0)log_2p(0) - p(1)log_2p(1)$, where $p(0)$ ($p(1)$) is the probability of the pixels being $0$ ($1$).

To reflect the difference of local patterns of the images,
we use a finite sized window $W^{x,y}_{a,b}$ to define:
\begin{equation}
T^{\mathcal{W}^{x,y}_{a,b}} =  \{T^{\mathcal{AD}}_{m=x,\cdots,x+a-1; n=y,\cdots,y+b-1}\},
\label{eq::capture}
\end{equation}
where $a$ and $b$ are the length and width of the window respectively, and $T^{\mathcal{W}^{x,y}_{a,b}}_{1,1} = T^{\mathcal{AD}}_{x,y}$.
We use the window $W^{x,y}_{a,b}$ to scan the binary image $T^{\mathcal{AD}}$ by moving one site for each step,
so the total number of $T^{\mathcal{W}^{x,y}_{a,b}}$ is $(\chi_1-a+1)(\chi_2-b+1)$. 
We get the cumulative Shannon entropy $S_{T^{\mathcal{AD}}}$ from binary image $T^{\mathcal{AD}}$ by:
\bea
&&S_{T^{\mathcal{AD}}} = \frac{\sum_{x=1}^{\chi_1-a+1}\sum_{y=1}^{\chi_2-b+1}S(T^{\mathcal{W}^{x,y}_{a,b}})}{(\chi_1-a+1)(\chi_2-b+1)}, \\
\nonumber \\
&&S(T^{W^{x,y}_{a,b}}) = -\sum_{\emph{pix}=\{0,1\}} P_{W^{x,y}_{a,b}}(\emph{pix})log_2P_{W^{x,y}_{a,b}}(\emph{pix}),
\label{eq:windows_entropy}
\eea
where $S(T^{W^{x,y}_{a,b}})$ is the Shannon entropy of $T^{\mathcal{W}^{x,y}_{a,b}}$, and \emph{pix} represents the pixel value.
The probability of $\emph{pix}$ ($\emph{pix}=0$ or $1$) $P_{W^{x,y}_{a,b}}(\emph{pix})$ is $\frac{n_{\emph{pix}}}{ab}$
with $n_{\emph{pix}}$ the number of $\emph{pix}$ in $T^{\mathcal{W}^{x,y}_{a,b}}$.

As $T^{\mathcal{AD}}$ only includes two pixels that are $0$ and $1$, we can intuitively view $T^{\mathcal{AD}}$
as the classical spin configurations $\tilde{T}^{\mathcal{AD}}$, where pixels $0$ ($1$) corresponds to spin up (down).
Choosing a finite region of $\tilde{T}^{\mathcal{AD}}$ by the window $W_{x,y}^{a,b}$,
we get a spin configuration $\tilde{T}^{\mathcal{W}^{x,y}_{a,b}}$, and we interpret it as a product state
$|\tilde{T}^{\mathcal{W}^{x,y}_{a,b}}\rangle$.
From this perspective, we define a quantity named virtual correlation $E_{T^{\mathcal{W}}}$:
\begin{equation}\label{eq::energy_image}
  E_{T^{\mathcal{W}}} =  -\frac{\langle \tilde{T}^{\mathcal{W}^{x,y}_{a,b}}| \sum_{<i,j>}{\sigma_i^z \sigma_j^z}
   |\tilde{T}^{\mathcal{W}^{x,y}_{a,b}}\rangle}
  {ab \langle \tilde{T}^{\mathcal{W}^{x,y}_{a,b}} | \tilde{T}^{\mathcal{W}^{x,y}_{a,b}}\rangle },
\end{equation}
where $\sigma^z$ is Pauli operator, $<i,j>$ means $i$ and $j$ are nearest neighbours.
\\

\section{Results}
We firstly examine VCS on the 1D transverse field Ising model (TFIM) \cite{sachdev2007quantum,LSM61exact} with the Hamiltonian $\hat{H} = \sum_{i}\hat{S}_{i}^{z}\hat{S}_{i+1}^{z} - h_x\sum_{i}\hat{S}_{i}^{x}$,
where $\hat{S}_{i}^{z}$ and $\hat{S}_{i}^{x}$ are the z- and x-component spin operators, respectively, and
$h_x$ is the transverse field.
With different $h_x$, a Landau-type \cite{landau1937theory} quantum phase transition occurs at the quantum critical point (QCP) $h_x =0.5$, which separates the antiferromagnetic (AFM) and paramagnetic (PM) phases.
Using DMRG \cite{white1992density, schollwock2011density} to calculate the
ground states for different transverse fields, we write the ground states
in the form of matrix product states (MPS) \cite{schollwock2011density,VC06MPSFaithfully}:
\begin{equation}\label{eq::mps}
  |\Psi\rangle = \sum_{\{s\}} A^{[1]}\cdots A^{[s_{i-1}]}\Phi^{[s_i]} B^{[s_{i+1}]}\cdots B^{[s_L]},
\end{equation}
where $|\Psi\rangle$ is in the mixed canonical form \cite{schollwock2011density},
$A^{[s_i]} = \{ A^{d}_{m,n} \}$ and $ B^{[s_i]} = \{B^{d}_{m,n}\}$
are in the left- and right-orthogonal forms, i.e. $\sum_{s_i}(A^{[s_i]})^{\dag} A^{[s_i]} = \sum_{s_i}B^{s_i}(B^{s_i})^\dag = \mathbb{I}$.
$\Phi^{[s_i]} = \{ \Phi^{d=1,2}_{m=1,\cdots,\chi;n=1,\cdots,\chi} \}$ is the canonical central
tensor~\cite{schollwock2011density}.
In our DMRG calculation, we take the system size $L = 80$, and dimension cut-off $\chi=30$

We treat the central tensor $\Phi$ as the feature of ground state $|\Psi\rangle$. As each element of $\Phi^{d}_{m,n}$ ranges from $-1$ to $1$ and the dimension $\sigma$ of the index $d$ of $\Phi$ equals to $2$, $\Phi$ cannot be directly interpreted as a color image.
The map $\mathcal{G}$ for $\Phi \overset{\mathcal{G}} {\to} \Phi^{\mathcal{G}}$ is as
\begin{equation}
{\Phi}^{\mathcal{G},d} =\left\{
\begin{array}{rcl}
\frac{1 + \Phi^{d}}{2},  &      & d = 1,2 \\
\\
0,                       &      & d = 3 \\
\end{array} \right.
\label{eq::G}
\end{equation}
we thus obtain the colorful image interpretations of the ground states with different transverse fields $h_x$ as shown in Fig.~\ref{fig::Ising_image} (a),
where $\Phi^{\mathcal{G}}[h_x=0.2]$, $\Phi^{\mathcal{G}}[h_x=0.5]$,
and $\Phi^{\mathcal{G}}[h_x=0.8]$ represent the AFM phase, QCP, and PM phase, respectively.

Unlike the image $T^{\mathcal{G}}[cat]$ in Fig.~\ref{fig::cat_image} (a) that directly shows the visual sense of a cat,
the image $\Phi^{\mathcal{G}}[h_x = 0.2 ~(0.5, 0.8)]$ does not show a distinguished visual information of different
quantum phases. Most of the pixels of $\Phi^{\mathcal{G}}[h_x = 0.2 ~(0.5, 0.8)]$ are green except a few distinguished pixels marked by red circles.
For the colorful images $\Phi^{\mathcal{G}}[h_x = 0.2 ~(0.5, 0.8)]$, the quantum phase transition of TFIM is hard to specify.
We show the histogram of tensor elements of $\Phi$ with $h_x$ = $0.2$, $0.5$, and $0.8$ in Fig.~\ref{fig::Ising_image} (d),
which indicates that the values of the most tensor elements are at the range of $-10^{-4}$ to $10^{-4}$.
The distribution of the count occurrence of tensor elements does not changed distinguishably with the various of $h_x$.
This explains why it is hard to classify the quantum phase transitions from the color images $\Phi^{\mathcal{G}}$ as shown in Fig.~\ref{fig::Ising_image} (a).

Choosing the threshold values $\alpha^{d=1} = \alpha^{d=2} = 0.5$ to segment $\Phi^{\mathcal{G}}[h_x = 0.2 ~ (0.5, 0.8)]$ by Eq.~(\ref{eq::B}), we obtain the binary images $\Phi^{\mathcal{B}}[h_x = 0.2 ~ (0.5,0.8)]$ as shown in Fig.~\ref{fig::Ising_image} (b).
This step is to obtain the binary images $\Phi^{\mathcal{B}}$ according to the sign of tensor elements of $\Phi$.
The pattern given by $\Phi^{\mathcal{B}}[h_x = 0.2, ~(0.5, 0.8)]$ looks randon, from which the quantum phase transition
AFM $\rightarrow$ PM cannot be characterized by the visual sense.

Based on the binary images $\Phi^{\mathcal{B}}[h_x = 0.2,~(0.5,0.8)]$,
we obtain the new binary image $\Phi^{\mathcal{AD},\delta_{1,2}}[h_x = 0.2, ~(0.5,0.8)]$ by Eq.~(\ref{eq::AD}) as shown in Fig.~\ref{fig::Ising_image} (c).
It is obviously that the pattern of $\Phi^{\mathcal{AD},\delta_{1,2}}[h_x=0.2]$ and $\Phi^{\mathcal{AD},\delta_{1,2}}[h_x=0.8]$
show a distinguishable difference.
The binary pixels in $\Phi^{\mathcal{AD},\delta_{1,2}}[h_x=0.2]$ are non-regular distributed while it presents a
regular configuration like a brick wall in $\Phi^{\mathcal{AD},\delta_{1,2}}[h_x=0.8]$.
More interestingly, at the critical transverse field $h_x=0.5$, the corresponding binary image $\Phi^{\mathcal{AD},\delta_{1,2}}[h_x=0.5]$ shows a mixed pattern of $\Phi^{\mathcal{AD},\delta_{1,2}}[h_x=0.2] $
and $\Phi^{\mathcal{AD},\delta_{1,2}}[h_x=0.8]$.
As shown in Fig.~\ref{fig::Ising_image} (c), the upper-left part (marked by a red triangular) of $\Phi^{\mathcal{AD},\delta_{1,2}}[h_x=0.5]$ is
a regular configuration like brick wall, and the pixels in lower-right part of $\Phi^{\mathcal{AD},\delta_{1,2}}[h_x=0.5]$
are non-regular distributed like the pixels in $\Phi^{\mathcal{AD},\delta_{1,2}}[h_x=0.2]$.

From the binary images $\Phi^{\mathcal{AD},\delta_{1,2}}[h_x=0.2,~(0.5,0.8)]$, the quantum phase transition from AFM to PM
of TFIM is visualized, and the phase transition can be characterized by the variation of the pattern
in $\Phi^{\mathcal{AD},\delta_{1,2}}[h_x=0.2,~(0.5,0.8)]$.
More binary images $\Phi^{\mathcal{AD},\delta_{1,2}}$ with different system sizes $L$, and dimension
cut-off $\chi$ are in the {\bf{Appendix}}.

\begin{figure}[htb]
\includegraphics[width=0.95 \columnwidth]{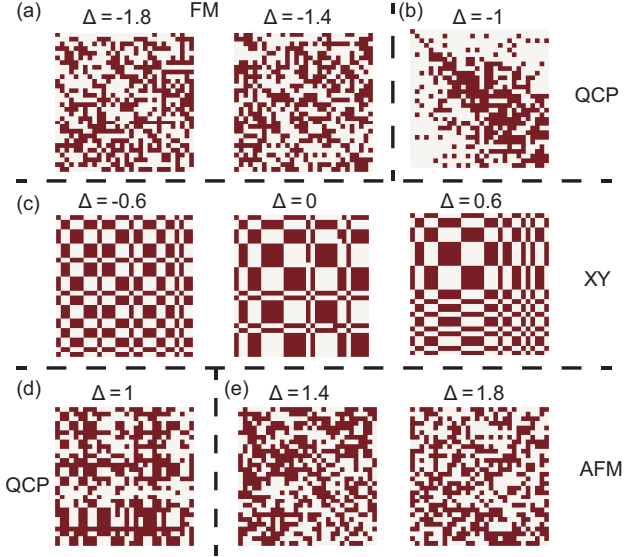}
\caption{(Color online) The binary images $\Theta^{\mathcal{AD}, \delta_{1,2}}$ of the ground states in the 1D anisotropic XXZ antiferromagnetic Heisenberg model (XXZHM) with anisotropy $\Delta$.
(a) $\Theta^{\mathcal{AD},\delta_{1,2}}$ in ferromagnetic (FM) phase with $\Delta=-1.8$ and $-1.4$. The textures are
in inhomogeneous patterns. (b) $\Theta^{\mathcal{AD},\delta_{1,2}}$ of the quantum critical point (QCP) $\Delta=-1$, where
FM and XY phases are separated. The texture shows a distorted pattern.
(c) $\Theta^{\mathcal{AD},\delta_{1,2}}$ in ferromagnetic (XY) phase with $\Delta=-0.6$, $0$ and $0.6$.
The textures show a homogeneous patterns like the regular brick wall and
display approximate spatial uniformity and consistency (global homogeneity).
(d) $\Theta^{\mathcal{AD},\delta_{1,2}}$ of the QCP $\Delta=1$, where
XY and antiferromagnetic (AFM) phases are separated. The texture is a mixture of
homogeneous patterns (the bottom of the image) and inhomogeneous patterns (the upper part of the image).
(e) $\Theta^{\mathcal{AD},\delta_{1,2}}$ in AFM phase with $\Delta=1.4$ and $1.8$.
The textures are in inhomogeneous patterns, where the distribution of the pixels is non-regularly.
For DMRG, we take the size of the system $L = 80$ and the dimension cut-off $\chi = 30$.}
\label{fig::xxzhalf_image}
\end{figure}

\begin{figure}[htb]
\includegraphics[width=0.95 \columnwidth]{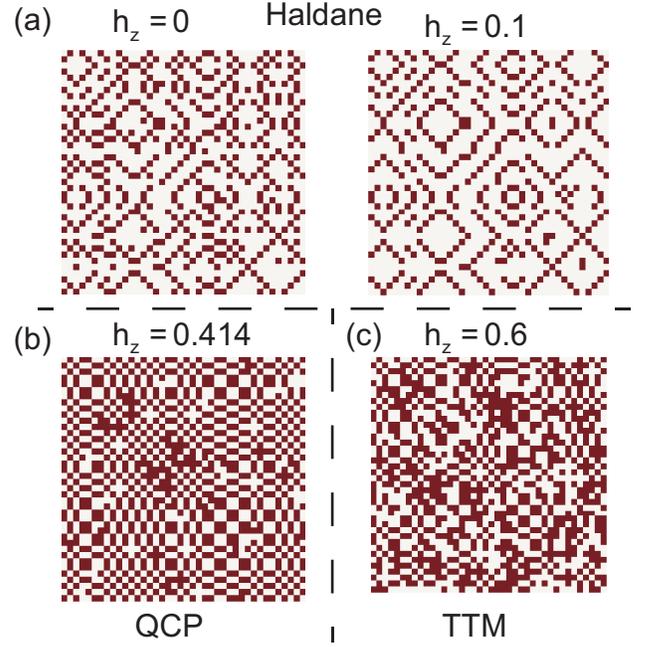}
\caption{(Color online) The binary images $\Omega^{\mathcal{AD}, \delta_{1,3}}$ of ground states in the spin-$1$
Heisenberg antiferromagnetic model with different magnetic field $h_z$.
(a) $\Omega^{\mathcal{AD},\delta_{1,3}}$ in Haldane phase with $h_z=0$ and $0.1$. The textures are
homogeneous and show the regularity of the lattice-window with diamond holes.
(b) $\Omega^{\mathcal{AD},\delta_{1,3}}$ of quantum critical point (QCP) $h_z=0.414$. The texture is like
a weak-distorted brick wall.
(c) $\Omega^{\mathcal{AD},\delta_{1,3}}$ in topological trivial magnetic phase (TTM) with $h_z=0.6$.
The texture are inhomogeneous.
For DMRG, we take the size of the system $L = 120$ and the dimension cut-off $\chi = 40$.}
\label{fig::haldane_image}
\end{figure}

To further demonstrate the validity of characterizing quantum phase transitions by VCS, we
turn to the 1D spin-$\frac{1}{2}$ anisotropic XXZ Heisenberg model (XXZHM) $H = \sum_{i}(S_i^xS_{i+1}^x + S_i^yS_{i+1}^y)
+ \Delta \sum_{i}S_i^zS_{i+1}^z$ \cite{giamarchi2003quantum} with $\Delta$ the magnetic anisotropy.
The system goes through three phases, which are the FM ($\Delta < -1$), XY ($-1 < \Delta < 1$), and AFM ($\Delta > 1$) phases.
The canonical central tensor of the ground states is $\Theta$. According to VCS,
the binary images $\Theta^{\mathcal{AD},\delta_{1,2}}$ is obtained by the procedure:
$\Theta \xrightarrow[Eq.~(\ref{eq::G})]{\mathcal{G}} \Theta^{\mathcal{G}}  \xrightarrow[Eq.~(\ref{eq::B})]{\mathcal{B}}
\Theta^{\mathcal{B}} \xrightarrow[Eq.~(\ref{eq::AD})]{\mathcal{AD}} \Theta^{\mathcal{AD},\delta_{1,2}} $,
where the threshold value for the binarization $\mathcal{B}$ is $\alpha^{d=1}=\alpha^{d=2}=0.5$.

Fig.~\ref{fig::xxzhalf_image} shows the binary images $\Theta^{\mathcal{AD},\delta_{1,2}}$
in AFM, XY, and FM phases.
When the system is in AFM and FM phase, the textures of the $\Theta^{\mathcal{AD},\delta_{1,2}}$
are non-regularly, as shown in Fig.~\ref{fig::xxzhalf_image} (a) and (c).
However, when the system is in XY phase, the textures of the $\Theta^{\mathcal{AD},\delta_{1,2}}$ are regularly distributed like brick walls,
as shown in Fig.~\ref{fig::xxzhalf_image} (b).
The phase transition points can be accurately identified as the abrupt change of the image style of $\Theta^{\mathcal{AD},\delta_{1,2}}$, which happens at $\Delta=1$ and $\Delta=-1$, respectively.
More binary images $\Theta^{\mathcal{AD},\delta_{1,2}}$ with different system sizes $L$ and dimension cut-off $\chi$ are in the {\bf{Appendix}}.

Determining the topological phases is a challenging task in quantum many-body systems \cite{wen1989vacuum, wen1990ground,gu2009tensor}.
Considering the spin-1 Heisenberg model (spin-1 HM) in a magnetic field with the Hamiltonian $H = \sum_{i}\sum_{\gamma=x,y,z} S_i^{\gamma}S_{i+1}^{\gamma} - h_z\sum_{i}S_i^{z}$, we use the VCS to visualize its ground states.
For $h_z < 0.414$, the system is in a symmetry protect topological phase known as Haldane phase \cite{haldane1983continuum,haldane1983nonlinear,white1993numerical} with
non-trivial boundary excitations and long-range string orders \cite{nijs1989preroughening,anfuso2007fragility}.
For $h_z > 0.414$, the system enters a topologically trivial magnetic (TTM) phase \cite{gu2009tensor}.
The canonical central tensor of the ground states is $\Omega=\{\Omega^{d=0,\cdots,d_{max}}_{l=0,\cdots,\chi;r=0,\cdots,\chi}\}$,
where $d_{max} =3$, and $\chi=40$.
We obtain the binary image $\Omega^{\mathcal{AD},\delta_{1,3}}$ by the VCS scheme:
$\Omega \xrightarrow {\mathcal{G}'} \Omega^{\mathcal{G}'}  \xrightarrow[Eq.~(\ref{eq::B})]{\mathcal{B}}
\Omega^{\mathcal{B}} \xrightarrow[Eq.~(\ref{eq::AD})]{\mathcal{AD}} \Omega^{\mathcal{AD},\delta_{1,3}} $, where
$\Omega^{\mathcal{G}'}$ = $(1+\Omega)/2$ and the threshold value for the binarization $\mathcal{B}$ is $\alpha^{d=1}=\alpha^{d=2}=0.5$.

Fig.~\ref{fig::haldane_image} (a) shows the binary images $\Omega^{\mathcal{AD},\delta_{1,3}}$ in Haldane phases with
$h_z = 0$ and $h_z=0.1$, where a regular pattern like the lattice-window with diamond holes is displayed.
Although a few parts exhibits local non-homogeneity, the whole image shows approximate spatial uniformity and periodicity.
The pattern given by $\Omega^{\mathcal{AD},\delta_{1,3}}$ in Haldane phase also exhibit self-similarity.
Fig.~\ref{fig::haldane_image} (b) shows the $\Omega^{\mathcal{AD},\delta_{1,3}}$ at the quantum critical point (QCP) with
$h_z=0.414$, where the image gives a weak distorted brick wall pattern.
The spatial arrangements of the texture become more dense.
When the system is in TTM with $h_z=0.6$ as shown in Fig.~\ref{fig::haldane_image} (c), the homogeneous pattern of
$\Omega^{\mathcal{AD},\delta_{1,3}}$ is absent.
Phase transition from Haldnae to TTM corresponds to the changes of patterns of $\Omega^{\mathcal{AD},\delta_{1,2}}$ and $\Omega^{\mathcal{AD},\delta_{2,3}}$, as shown in Fig.A4 in {\bf{Appendix}}.
The specific self-similarity of Haldane phase supports that topological phases can also be characterized by VCS. \\

\begin{figure*}[htb]
\includegraphics[width=1.6 \columnwidth]{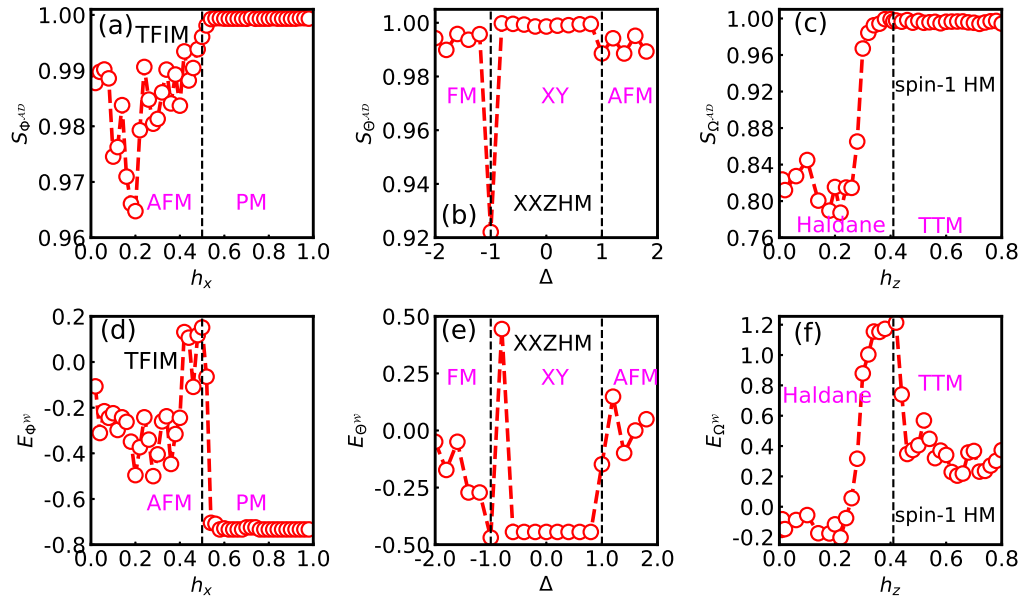}
\caption{(Color online) Cumulative Shannon entropies $S_{\Phi^{AD}}$, $S_{\Theta^{AD}}$ and $S_{\Omega^{AD}}$ of binary images $\Phi^{AD}$, $\Theta^{AD}$ and $\Omega^{AD}$ are shown in (a), (b) and (c), respectively, and the corresponding virtual correlations
$E_{\Phi^{\mathcal{W}}}$, $E_{\Theta^{\mathcal{W}}}$ and $E_{\Omega^{\mathcal{W}}}$ are shown in (d), (e) and (f).
The image sizes of $S_{\Phi^{AD}}$, $S_{\Theta^{AD}}$ and $S_{\Omega^{AD}}$ are $\chi\times\chi = 30\times 30$, $30\times 30$ and
$40 \times 40$.
The windows for getting the cumulative Shannon entropies $S_{\Phi^{AD}}$, $S_{\Theta^{AD}}$ and $S_{\Omega^{AD}}$
are $W^{x_1,y_1}_{10,10}$, $W^{x_2,y_2}_{10,10}$ and $W^{x_3,y_3}_{10,10}$, where $1\leq x_1 ( y_1, x_2, y_2) \leq 21$ and
$1\leq x_3 (y_3) \leq 31$.
The windows for getting the virtual correlations $E_{\Phi^{\mathcal{W}}}$, $E_{\Theta^{\mathcal{W}}}$, $E_{\Omega^{\mathcal{W}}}$
are $W^{1,1}_{30,30}$, $W^{1,1}_{9,9}$, and $W^{1,1}_{40,40}$, respectively.
In DMRG, the system sizes for the phases of TFIM, XXZHM and spin-1 HM are taken as L=80, 80 and 200, respectively.
}
\label{fig::EE3model}
\end{figure*}

From the perspective of visualization, we note that phase transitions correspond to the drastic change
of the binary patterns of $\Phi^{\mathcal{AD}}$, $\Theta^{\mathcal{AD}}$ and $\Omega^{\mathcal{AD}}$.
Here, we use two quantities namely the cumulative Shannon entropies and virtual correlations
of binary images to quantitatively characterize the phase transitions.

As shown in Figs.~\ref{fig::EE3model} (a), (b) and (c), the cumulative Shannon entropies $S_{\Phi^{\mathcal{AD}}}$,
$S_{\Theta^{\mathcal{AD}}}$ and $S_{\Omega^{\mathcal{AD}}}$ in different quantum phases display different behaviours.
For TFIM, when the system is in the magnetic ordered phases, $S_{\Phi^{\mathcal{AD}}}$ shows an oscillatory behaviour, which means the local patterns of $\Phi^{\mathcal{AD}}$ are unstable with different transverse field $h_x$. However, when the system is in disordered paramagnetic phase, $S_{\Phi^{\mathcal{AD}}}$ is stable and keeps a constant value.
The behaviour of $S_{\Theta^{\mathcal{AD}}}$ versus $\Delta$ for XXZHM keeps consist with the behaviour of $S_{\Phi^{\mathcal{AD}}}$ versus $h_x$, which implies that $S_{\Theta^{\mathcal{AD}}}$ is oscillating in magnetic ordered AFM (FM) phase and converged in magnetic disordered XY phase.
For the system with topological properties such as the spin-1 HM,
the cumulative Shannon entropies $S_{\Omega^{\mathcal{AD}}}$ reaches the saturated value when the system is in TTM phase.

By means of the cumulative Shannon entropies, the amount of information given by the distribution of binary pixels are measured.
The cumulative Shannon entropy is an indicator that explains how much randomness of the binary pixels is.
The more uniform of the distribution of the random binary pixels is, the more the cumulative Shannon entropies will be.
For the phases of TFIM, XXZHM, and spin-1 HM, the textures of $\Phi^{\mathcal{AD}}$, $\Theta^{\mathcal{AD}}$ and $\Omega^{\mathcal{AD}}$
in PM phase, XY phase and TTM phase are more uniform, so the
corresponding values of cumulative Shannon entropy are larger.

The virtual correlations $E_{\Phi^{\mathcal{W}}}$, $E_{\Theta^{\mathcal{W}}}$, and $E_{\Omega^{\mathcal{W}}}$  of binary images
$\Phi^{\mathcal{AD}}$, $\Theta^{\mathcal{AD}}$ and $\Omega^{\mathcal{AD}}$ are shown in Fig.~\ref{fig::EE3model}(d), (e) and (f),
respectively. Obviously, virtual correlations can also reflect the information of quantum phase transitions.

\emph{Conclusion}---In summary, we have developed an image segmentation scheme named VCS
to visualize and characterize the quantum phases and phase transitions for many-body systems.
There are three steps in VCS: first interpreting the canonical central tensor of the ground states
into a color image by mapping $\mathcal{G}$, and second binarizing the color image in each color channel by mapping $\mathcal{B}$, then third calculating the absolute differences of the binary images related to color channels by mapping $\mathcal{AD}$.
The colorful image interpretation of the ground states by mapping $G$ cannot give an
effective information of renormalized quantum states and quantum phase transition due to the large variance of tensor elements.
The selection of threshold in $\mathcal{B}$ is the key in our scheme,
and a wrong selection of the threshold value may misinterpret the background pixel and result in a
degradation of scheme performance.
For the ground states of quantum many-body systems, we chose the threshold value as $0.5$, which equals to binarizing the canonical central tensor of the ground states based on signs of tensor elements.

Based on VCS, the renormalized quantum states represented by binary images show different patterns in distinct quantum phases, from which the quantum critical points can be identified.
The success of VCS is demonstrated on three 1D quantum many-body spin models that possess various quantum
phases, including the conventional phases within Landau paradigm and the topological phases with nonlocal orders.
When the TFIM is in PM phase, and the spin-$\frac{1}{2}$ XXZHM is in XY phase,
the corresponding binary image patterns show a specific brick wall pattern, while the lattice-window with diamond holes
emerges in the texture pattern of the spin-1 Heisenberg model in Haldane phase.
These patterns display the spatial homogeneity, self-similarity, periodicity, and directionality, which may
be related to the underlying physical properties of the quantum phases. When the systems are at the QCP, the
image pattern is a mixture of the patterns of the two adjacent quantum phases.

Our scheme gives a perceptual analysis of the renormalized quantum states in many-body systems based on the image segmentation.
Other techniques in computer vision such as edge detection \cite{sundani2015edge}, digital image processing  \cite{gonzalez2008digital} may also be utilized in this scheme to obtain a better performance of the image interpretation of quantum states.
This would motivate people to seek for techniques in computer vision to unveil more novel properties of quantum
states of many-body systems.

\begin{acknowledgments}
This work is supported in part by the NSFC (Grant No. 11834014), the National Key R\&D Program of China (Grant No. 2018FYA0305804),  the Strategetic Priority Research Program of the Chinese Academy of Sciences (Grant No. XDB28000000), and Beijing Municipal Science and Technology Commission (Grant No. Z118100004218001).
SJR is supported by Beijing Natural Science Foundation (No. 1192005 and No. Z180013), Foundation of Beijing Education Committees (No. KM202010028013), and the Academy for Multidisciplinary Studies, Capital Normal University.
\end{acknowledgments}

\setcounter{equation}{0}
	\setcounter{figure}{0}
	\setcounter{table}{0}
	\makeatletter
	\renewcommand{\theequation}{A\arabic{equation}}
	\renewcommand{\thefigure}{A\arabic{figure}}
	\renewcommand{\thetable}{A\arabic{table}}
\begin{appendix}

\section{Binary images representation of quantum states with different system sizes}

\begin{figure*}[htb]
	\includegraphics[width=0.8\linewidth]{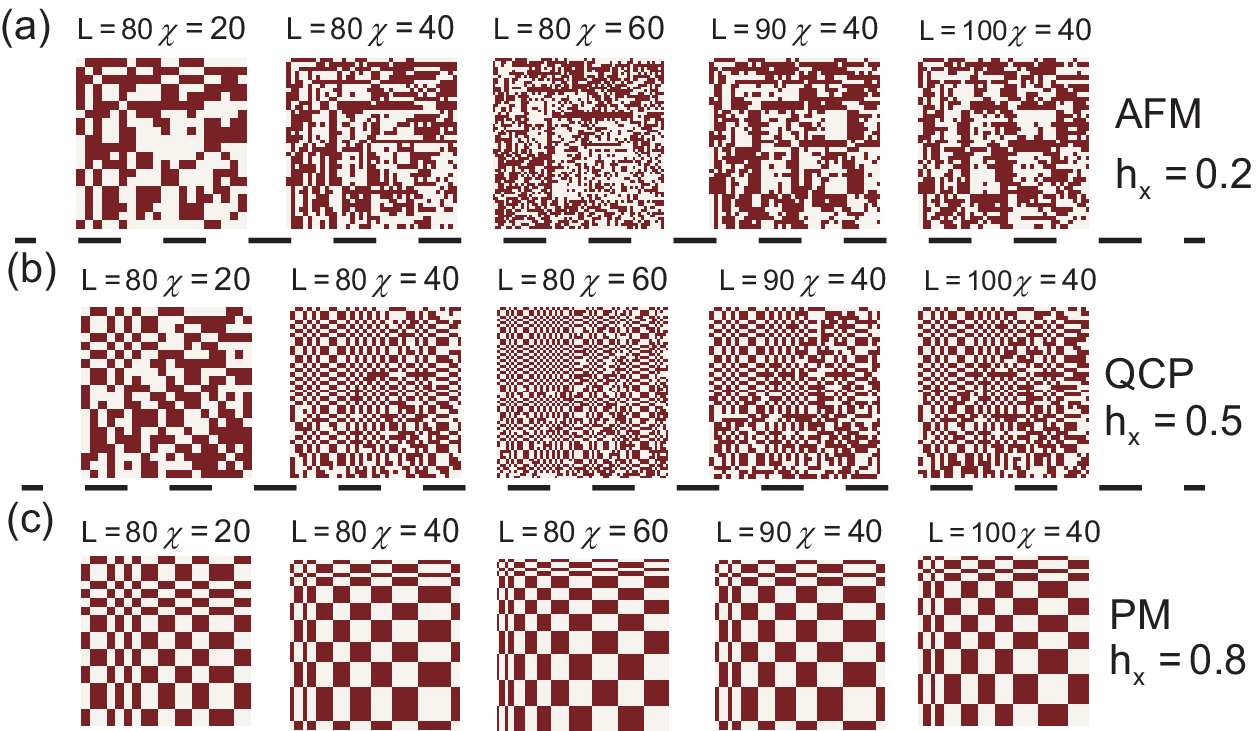}
	\caption{(Color online) Binary images $\Phi^{\mathcal{AD},\delta_{1,2}}$ are in the antiferromagnetic (AFM) phase with
transverse field $h_x=0.2$ (a) and quantum paramagnetic (PM) phase with $h_x=0.8$ (c), and at the quantum critical point (QCP) $h_x=0.5$ (b). In FM and AFM phases and at the QCP, the textures of $\Phi^{\mathcal{AD},\delta_{1,2}}$ with different dimension cut-off $\chi=20$, $40$, $60$ and different system size $L=80$, $90$, $100$ are presented.
The size of the binary images $\Phi^{\mathcal{AD},\delta_{1,2}}$ is $\chi\times\chi$.}
	\label{fig::Ising_scaling}
\end{figure*}
\begin{figure*}[htb]
	\includegraphics[width=0.8\linewidth]{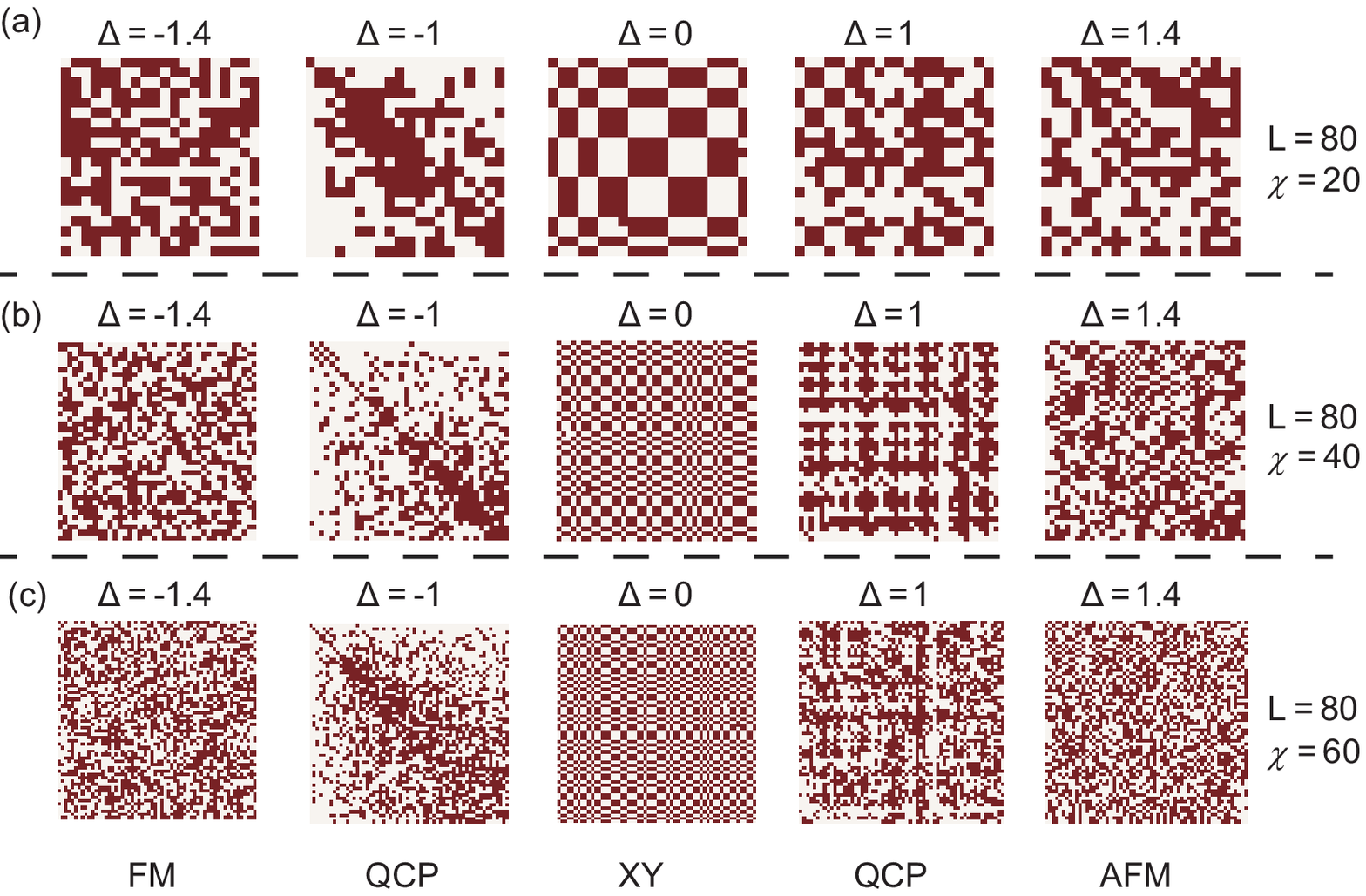}
	\caption{(Color online) Binary images $\Theta^{\mathcal{AD},\delta_{1,2}}$ with fixed system size $L=80$ and
different dimension cut-off $\chi=20$, $40$, and $60$ are given in (a), (b), and (c), respectively.
In each case of $L$ and $\chi$, $\Theta^{\mathcal{AD},\delta_{1,2}}$ with XXZ anisotropy $\Delta=-1.4$, $-1$, $0$, $1$, $1.4$ are presented, which correspond to FM phase, QCP (FM to XY), XY phase, QCP (XY to AFM), and AFM phase, respectively.
The size of the binary images $\Theta^{\mathcal{AD},\delta_{1,2}}$ is $\chi\times\chi$.}
	\label{fig::XXZ_scaling}
\end{figure*}

\begin{figure*}[htb]
	\includegraphics[width=0.8\linewidth]{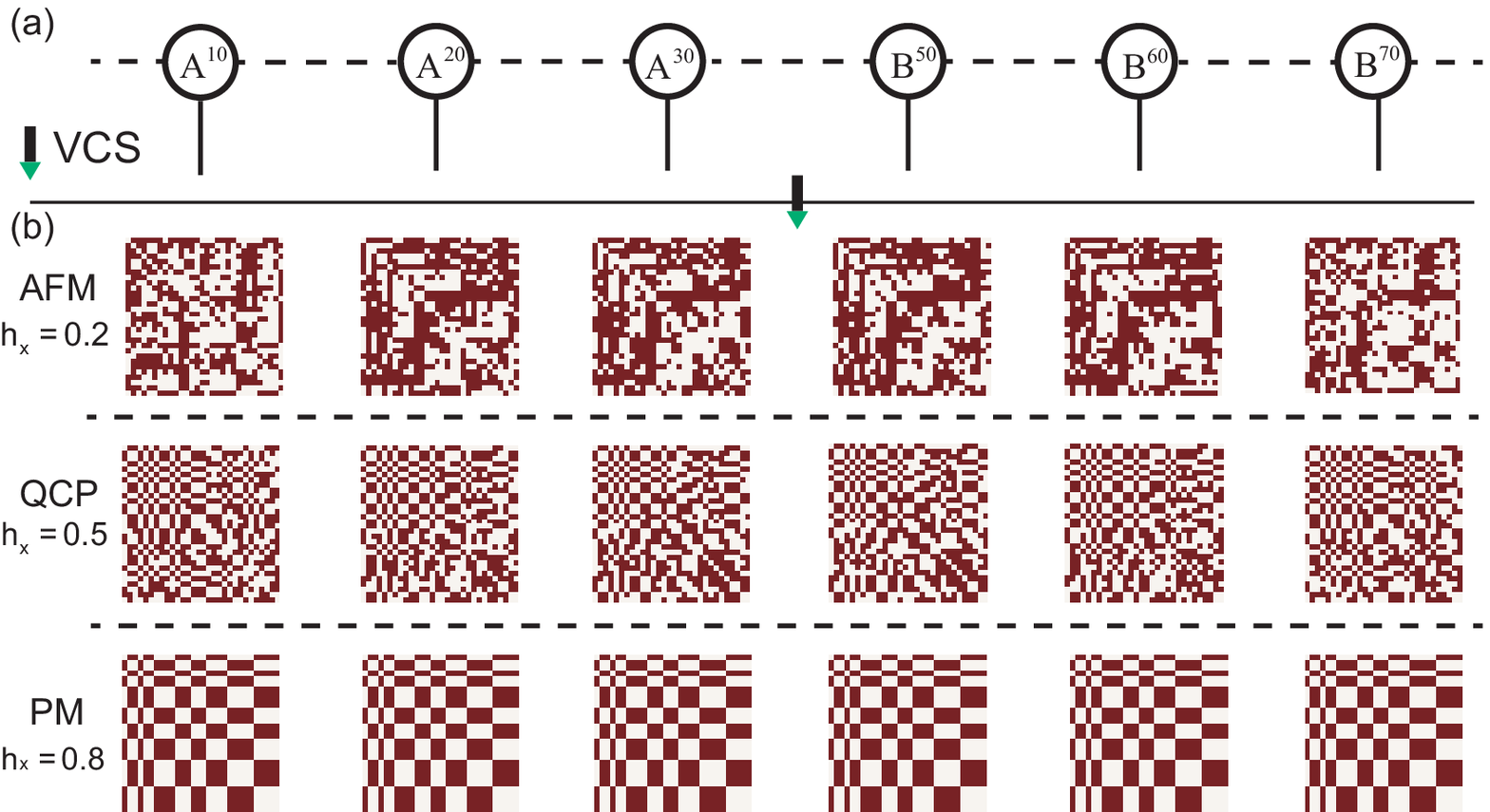}
	\caption{(Color online) (a) The ground states $\psi\rangle$ of the transverse field Ising model (TFIM) are represented in form of matrix product states. $\psi\rangle$ is in the mixed canonical form, where $A^{i}$ and $B^{j}$ are left- and right- normalized.
(b) By the scheme of VCS, $\{ A^{i}\}$ and $\{B^{j}\}$ are visualized by binary images with $i=10$, $20$, $30$
and $j=50$, $60$, $70$, where $i~(j)$ is the position of local tensor on MPS. Three cases for the TFIM in AFM phase ($h_x=0.2$), PM phase ($h_x=0.8$) and at the QCP ($h_x=0.5$) are presented. For DMRG, we take the size of system $L=80$ and
dimension cut-off $\chi=30$.
The size of binary images of $\{ A^{i}\}$ and $\{B^{j}\}$ is $\chi\times\chi$. }
	\label{fig::Ising_position}
\end{figure*}

\begin{figure*}[tbp]
	\includegraphics[width=0.8\linewidth]{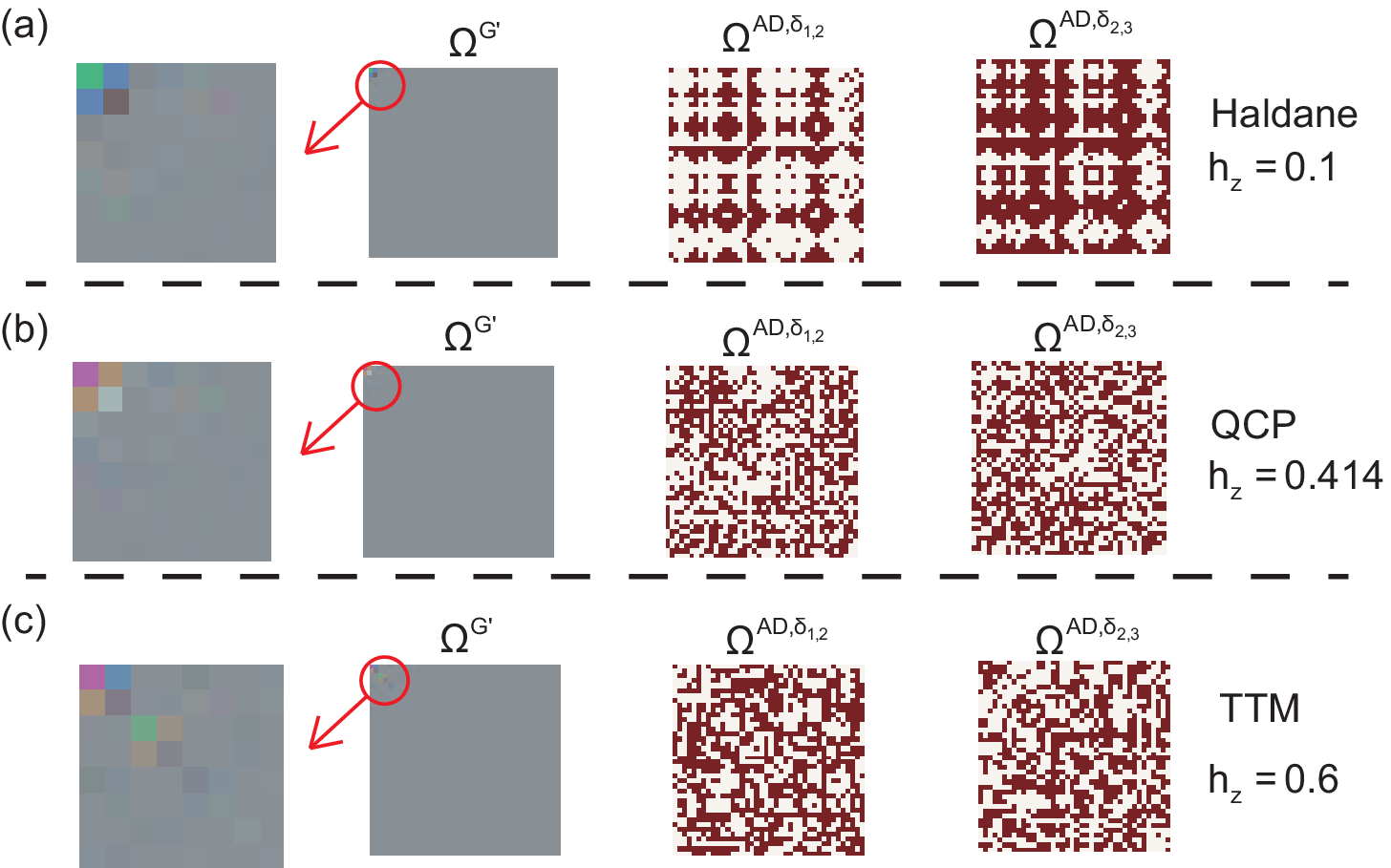}
	\caption{(Color online) The ground states of the spin-1 antiferromagnetic Heisenberg model in Haldane phase ($h_z=0.1$), topological trivial magnetic (TTM) phase ($h_z=0.6$), and at the quantum critical point (QCP) ($h_z=0.414$) are
visualized by color image $\Omega^{\mathcal{G}'}$ and binary image $\Omega^{\mathcal{AD},\delta_{1,2}}$
($\Omega^{\mathcal{AD},\delta_{2,3}}$) based on the VCS.
The distinguishable details of $\Omega^{\mathcal{G}'}$ with different $h_z$ are marked by red circles.
The system size $L=120$, and dimension cut-off in DMRG $\chi=40$.
The sizes of color images $\Omega^{\mathcal{G}'}$ and binary images $\Omega^{\mathcal{AD},\delta_{1,2}}$
($\Omega^{\mathcal{AD},\delta_{2,3}}$) are $3\times\chi\times\chi$ and $\chi\times\chi$.}
	\label{fig::Haldane_color}
\end{figure*}

By our designed image segmentation scheme VCS,
the quantum states of many-body systems can be represented by binary images.
Below, we show the textures of binary image representation of quantum states with
different system size $L$ and dimension cut-off $\chi$ to illustrate that the
special binary textures of quantum states are not from the
finite-size effects.

As shown in Fig.~\ref{fig::Ising_scaling}, the ground states of the transverse field Ising model (TFIM) are represented by
binary images $\Phi^{\mathcal{AD},\delta_{1,2}}$, where the ground states are calculated by density matrix renormalization group
(DMRG) method with system size $L=80$, $90$, $100$ and dimension cut-off $\chi=20$, $40$, $60$.
The sizes of binary images are $\chi\times\chi$.
When the dimension cut-off $\chi$ is fixed at $40$, the textures of $\Phi^{\mathcal{AD},\delta_{1,2}}$ with different system sizes $L=80$, $90$, $100$ keep a consistency, which indicate the specific binary textures of $\Phi^{\mathcal{AD},\delta_{1,2}}$  are not from the finite-size effect.
Whenever the system is in antiferromagnetic (AFM) phase, quantum paramagnetic (PM) phase or at the quantum critical point (QCP),
the textures of $\Phi^{\mathcal{AD},\delta_{1,2}}$ show specific self-similarity with increasing $\chi$ from
$20$, $40$ to $60$ when system size $L$ is fixed at $80$.
Self-similarity is the fundamental character of fractal geometry \cite{mandelbrot1982fractal}.
As the binary images $\Phi^{\mathcal{AD},\delta_{1,2}}$ are directly related to the sign
of the local tensor elements in ground states, the self-similarity may be related to some latent physics of the
quantum mechanic wavefunctions.

In Fig.~\ref{fig::XXZ_scaling}, we show the binary images $\Theta^{\mathcal{AD},\delta_{1,2}}$, which is the interpretation of
the ground states of the spin-$\frac{1}{2}$ XXZ anisotropy Heisenberg chain model.
We fix the system size at $L=80$, and  show the various of the textures of $\Theta^{\mathcal{AD},\delta_{1,2}}$ with different dimension cut-off $\chi=20$, $40$, and $60$.
When the system is in some quantum phases, the textures of $\Theta^{\mathcal{AD},\delta_{1,2}}$ show a self-similarity with
the increase of image size $\chi \times \chi$. By taking the standard XY phase ($\Delta=0$) as an example, we may see that the brick-wall pattern
of the textures does not change with the increase of $\chi$.

\section{Binary images of other local tensors in matrix product states}
The ground states $|\Psi\rangle$ of the TFIM in the form of MPS is in the main text (Eq.~($7$)),
where $|\Psi\rangle = \sum_{\{s\}} A^{[1]}\cdots A^{[s_{i-1}]}\Phi^{[s_i]} B^{[s_{i+1}]}\cdots B^{[s_L]}$ and
$|\Psi\rangle$ is in the mixed canonical form. $\Phi$ is the canonical central tensor and $\{A^{i}\}$ ($\{B^{j}\}$)
are the left- (right-) normalized.  In the main text, we treat $\Phi$ as the feature of the ground states and
use its binary images $\Phi^{\mathcal{AD},\delta_{1,2}}$ to characterize quantum phase transitions based on scheme VCS.

Here, we treat other local tensors $\{A^{i}\}$ ($\{B^{j}$) as the feature of $|\Psi\rangle$, and use VCS
to obtain its binary images as shown in Fig.~\ref{fig::Ising_position}.
Although each local tensor is different, but its binary images $\{A^{i}\}^{\mathcal{AD},\delta_{1,2}}$
along the MPS chain show the same textures,
which may be related to a kind of symmetry of the ground states.

\section{color images and binary images of the spin-1 chain}
In the main text, we use the binary image $\Omega^{\mathcal{AD},\delta_{1,3}}$ to
show different quantum phases of the spin-1 antiferromagnetic chain model.
The dimension of physical bond $\Omega$ is $3$, for each $\Omega$ under the fixed magnetic field $h_z$,
we can get three binary images such as $\Omega^{\mathcal{AD},\delta_{1,2}}$, $\Omega^{\mathcal{AD},\delta_{1,3}}$, and
$\Omega^{\mathcal{AD},\delta_{2,3}}$.
Here, we show the color images $\Omega^{\mathcal{G}'}$ and binary images $\Omega^{\mathcal{AD},\delta_{1,2}}$ ($\Omega^{\mathcal{AD},\delta_{2,3}}$) with different $h_z$ in Fig.~\ref{fig::Haldane_color}.

The visual sense of $\Omega^{\mathcal{G}'}$ is the same as the color image $\Phi^{\mathcal{G}}$ of the TFIM in Fig.~3 of main text, where most of the pixels are gray except a few distinguishable pixels in parts marked by red circles,
from which the quantum phase transition from Haldane phase to topological trivial magnetic (TTM) phase can hardly be identified.
The texture of binary images $\Omega^{\mathcal{AD},\delta_{1,2}}$ ($\Omega^{\mathcal{AD},\delta_{2,3}}$) in Haldane phase
with $h_z=0.1$ shows the pattern of many diamond-shaped clusters.
As the textures of $\Omega^{\mathcal{AD},\delta_{1,3}}$ in Haldane phase are the lattice-window with diamond holes,
$\Omega^{\mathcal{AD},\delta_{1,3}}$ and $\Omega^{\mathcal{AD},\delta_{1,2}}$ ($\Omega^{\mathcal{AD},\delta_{2,3}}$)
display the complementary symmetry.
When the system is at the QCP ($h_z=0.414$), such specific textures are broken, and when the system is in the TTM phase,
the textures are non-regular.

\end{appendix}
\input{QSseg_arxiv.bbl}
\end{document}

%% file: QSseg_arxiv.bbl
%